\documentclass[twocolumn, prl]{revtex4}
\pdfoutput=1
\usepackage{graphicx}
\usepackage{epsfig}
\begin{document}
\title{Bosons in a double-well potential: Understanding the interplay between disorder and interaction in a simple model }
\author{Qi Zhou and S. Das Sarma}
\affiliation{Joint Quantum Institute and Condensed Matter Theory Center, Department
of Physics, University of Maryland, College Park, MD 20742}
\date{\today}
\begin{abstract}

We propose an exactly solvable model to reveal the physics of the interplay between interaction and disorder in bosonic systems. Considering interacting bosons in a double-well potential, in which disorder is mimicked by taking the energy level mismatch between the two wells to be randomly distributed, we find  ``two negatives make a positive" effect. While disorder or interaction by itself suppresses the phase coherence between the two wells, both together enhance the phase coherence. This model captures several striking features of the disordered Bose-Hubbard model found in recent numerical simulations. Results at finite temperatures may help explain why a recent experiment did not find any evidence for the enhancement of phase coherence in a disordered bosonic system. 

\end{abstract}
\maketitle

Physics of disorder in both bosonic and fermionic systems has been attracting great interest from physicists for decades. While the effect of disorder on non-interacting particles can be well described by Anderson localization\cite{Anderson}, understanding the interplay between interaction and disorder still remains challenging\cite{Review}. Recent developments in studying cold atoms provide new opportunities to reveal the nature of interacting quantum particles in the presence of disorder\cite{Reviewatoms}. In atomic systems, the strength of both interaction and disorder can be well controlled experimentally, unlike in solid state systems. There has been significant recent progress in studying disorder effects in both non-interacting and interacting atomic systems\cite{Aspect,Ingusico1, Ingusico2, Demarco1, Demarco2, Demarco3, Randy}.  

From the theoretical side, to understand even the spinless bosons in the presence of both interaction and disorder is a non-trivial problem.  Scaling analysis, renormalization group theory and sophisticated numerical simulations are often used\cite{JS, Scaling, RG, RG2, NT, RS, QMC, DMRG}. It is desirable in this context to have some exactly solvable models, which capture all the ingredients, such as interaction, disorder and finite temperature effects. In this letter, we propose and solve such a model, which makes it possible to reveal the underlying physics transparently. Our minimal model captures the interplay between interaction and disorder in bosonic systems, showing conclusively that weak disorder in the presence of interaction generically enhances the phase coherence of the system. 

The model we propose describes interacting bosons in a disordered double-well potential. The Hamiltonian can be written as 
\begin{equation}
H=-t(b^\dagger_{L}b_R+c.c)+\frac{U}{2}\sum_{\sigma}n_{\sigma}(n_\sigma-1)+\frac{\epsilon}{2}(n_R-n_L),\label{h}
\end{equation}
where $b^\dagger_L$($b_L$) and $b^\dagger_R$($b_R$) are the creation(annihilation) operators in the left and right well respectively, $\sigma=L,R$, $n_L=b^\dagger_Lb_L$($n_R=b^\dagger_Rb_R$) is the number operator in the left (right) well, $t$ is the tunneling amplitude between the two wells, $U>0$ is the onsite interaction,  and $\epsilon$ is the energy level mismatch between the two wells. A special case of this model ($\epsilon\equiv0$) has been studied both theoretically and in cold atom experiments\cite{doublewell1,doublewell2, Ho}. In our case, $\epsilon$ is randomly distributed according to a certain probability function $P(\epsilon) $, thus simulating disorder. In experiments, this can be realized by randomly tilting the double-well potential. The thermodynamic quantities will be the ensemble averaged values. In this paper, we will focus on the case $P(\epsilon)=1/(2\Delta), \epsilon\in[-\Delta, \Delta] $, where $\Delta$ characterizes the disorder strength. Other distribution functions for $\epsilon$ do not change the qualitative conclusion presented in this paper. Eq.(\ref{h}) can be viewed as a two site version of the extensively studied disordered Bose-Hubbard(BH) model with random onsite energies.

Despite the simplicity of this model, many interesting phenomena are found by solving this model with $N$ particles. An illuminating example is the case $\epsilon\equiv0 $. It has been shown that the ground state has a crossover from the coherent state to the Fock state when $U/t$ increases\cite{Ho}, which mimics the well-known phase transition in the thermodynamic limit from a superfluid condensate to a Mott insulator state in the full BH model. The physics for either the crossover or the phase transition between the phases is the same in the minimal model and the BH model, namely, interaction suppresses number fluctuations and spatial phase coherence. In the case where $\epsilon$ is randomly distributed, we will see that the minimal model qualitatively captures the physics of interacting bosons in a disordered potential for a number of fundamental questions, even though it can not answer some quantitative questions, such as where there is a direct transition from Mott insulator to superfluid.

Before we discuss the details, we briefly summarize the main qualitative questions that we are going to answer in this paper. {\bf Q1}: Whether disorder can enhance the phase coherence between the two wells,  defined by ${C}=\langle b^\dagger_Lb_R\rangle/N=\langle b^\dagger_Rb_L\rangle/N$, in some parameter regimes, in contrast to the intuition that disorder must always destroy the phase coherence? {\bf Q2}: Whether interaction can also enhance the phase coherence in the presence of disorder, though without disorder it is known that repulsive interaction usually suppresses phase coherence? {\bf Q3}: In the clean system, $\epsilon\equiv0$, the compressibility of the system vanishes when the phase coherence is destroyed by interaction. Is this still true in the disordered case? {\bf Q4}: How does the phase coherence between the two wells depend on the temperature? We will see that the answers to these four questions provide insight into some striking features of the quantum phase diagram of the disordered BH model. Our answers to these questions on the minimal model also shed light on a recent disagreement between the prediction of numerical simulations \cite{QMC, DMRG}and experimental observations\cite{Demarco3}. 

{\bf The answer to Q1}: To solve the disordered problem, we start from the case of $N$ particles in the double-well with a fixed $\epsilon$. We write the Schr\"odinger equation $H|\Psi\rangle=E|\Psi\rangle$ in the Fock space. Define $|l\rangle=|N_L,N_R\rangle=|\frac{N}{2}-l,\frac{N}{2}+l\rangle$, where $l=0,\pm 1, \pm 2,...\pm N/2$. For simplicity, we assume $N$ is a large even number. Expanding $|\Psi\rangle=\sum_l\psi_l|l\rangle$, we obtain 
\begin{equation}
(E-E_l)|l\rangle=-tM_{l,l+1}|l+1\rangle-tM_{l-1,l}|l-1\rangle,
\end{equation}
where $E_l=U l^2+\epsilon l+\frac{U}{2}(N^2/2-N)$, $M_{l,l+1}=M_{l+1,l}=\sqrt{N/2(N/2+1)-l(l+1)}$. The eigenenergies and eigenfunctions can be easily computed by exact diagonalization. At zero temperature, the phase coherence between the two wells can be characterized as $C_{\epsilon}=\frac{1}{N}\langle b^\dagger_Lb_R\rangle_\epsilon=\frac{1}{N}\sum_lM_{l,l+1}\psi_l^0\psi_{l+1}^0$, where $\psi_l^0$ is the ground state wave function in the Fock space,  $\langle {O} \rangle$ is the expectation value of the operator ${O}$ in the ground state, and the subscript implies a fixed $\epsilon$. The results for the disordered case can then be obtained by averaging those results for fixed $\epsilon$ according to the distribution function $P(\epsilon)$, 
\begin{equation}
\overline{\langle O\rangle}=\int_{-\Delta}^\Delta d\epsilon P(\epsilon){\langle O\rangle}_\epsilon.\label{ave}
\end{equation}

The results for $C_\epsilon$ for both noninteracting case and finite $U$ are shown in Fig.(1A).  The numerical results with the total particle number $N=100$ have been chosen to illustrate the physics, which does not depend on the exact value of the total particle number. When $U=0$, $C_\epsilon$ quickly decreases with increasing $\epsilon$, since noninteracting particles can only tunnel from one well to the other if the energy mismatch $\epsilon$ is smaller than the tunneling $t$.  The behavior of $C_\epsilon$ as a function of $\epsilon$ depends explicitly on $U$, as shown in Fig.1. The dependence of $C_\epsilon$ on $\epsilon$ becomes more extended with increasing $U$, though $C_\epsilon(\epsilon=0)$ decreases with increasing $U$. More importantly,  the curves become non-monotonic when the discretization of $E_l$ induced by interaction becomes significant for $U\gtrsim tN$. Local maxima  of $C_\epsilon$ emerge at certain values of $\epsilon$.  For those values $\epsilon^*=U( 1-2l^*)$, where $l^*$ is an integer number, $E_{l^*}=E_{l^*-1}$ is satisfied. In other words, the state $|N/2-l^*,N/2+l^*\rangle$ has the same energy with the one $|N/2-l^*+1,N/2+l^*-1\rangle$. There is no extra interaction energy cost for one particle tunneling from one well to the other. The tunneling will be enhanced at $\epsilon^*$, forming gently varied bumps on the curves for $C_\epsilon$. This non-monotonic behavior of $C_\epsilon$ becomes even more dramatic when $U$ increases. For large values of $U$, only in narrow regions near $\epsilon$, tunneling is strongly enhanced. As a result, resonance features emerge with sharp peaks located at $\epsilon^*$ sitting on the slowly decaying envelope of $C_\epsilon$.  In the large $U$ limit, $E_l\gg E_{l^*}=E_{l^*-1}$ if $l\neq l^*$ or $l^*-1$, a two level approximation can be made, and it is straightforward to show that $C_{\epsilon^*}\sim M_{l^*-1,l^*}$. $C_{\epsilon^*}$ decreases slowly as $\epsilon^*$ increases, as shown in Fig.(1A). 
\begin{figure}[tbp]
\begin{center}
\includegraphics[width=3.4in]{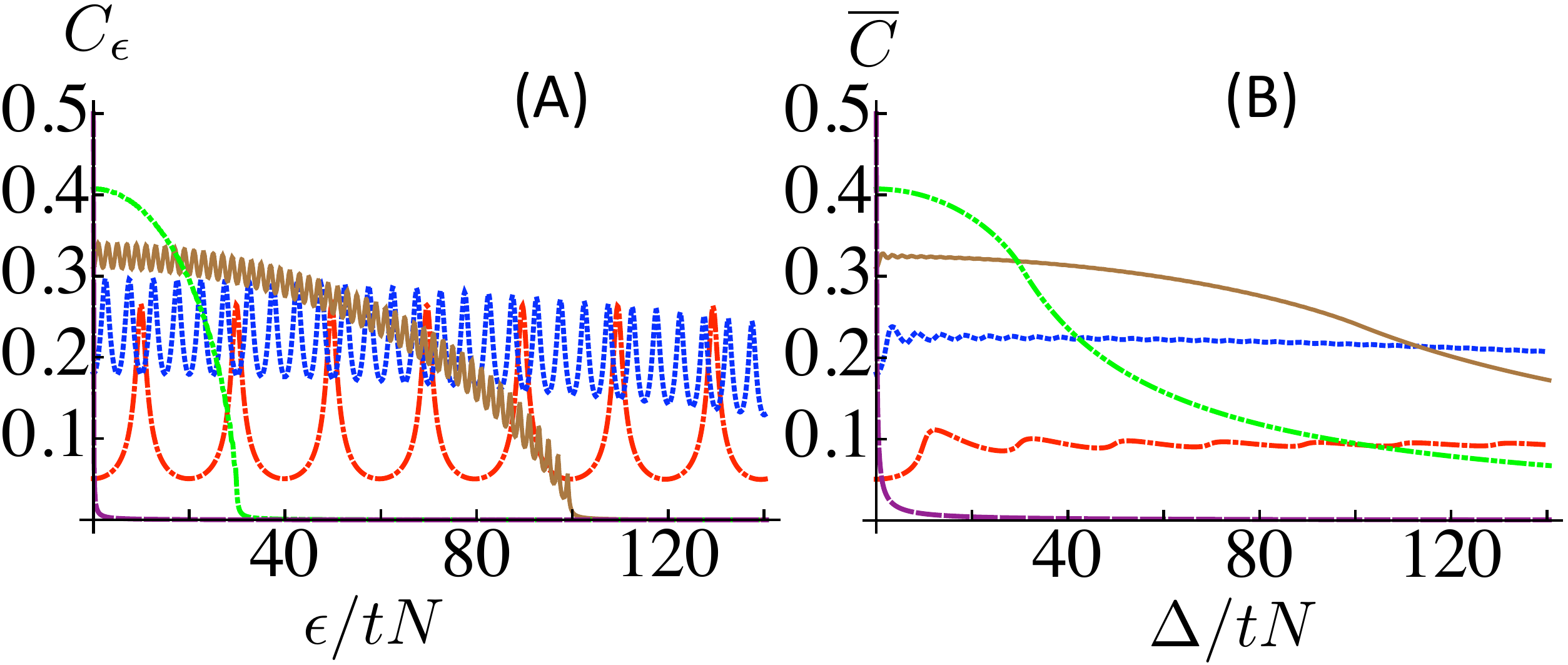}
\end{center}
\caption{A: $C_\epsilon$ as a function of $\epsilon/tN$ at different interaction. B: $\overline{C}$ as a function of $\Delta/tN$. Dashed(purple), dash-double-dotted(green), solid(brown), dotted(blue), dash-dotted(red) represent $U/tN=0$, $0.3$, $1$, $2.5$, $9.95$ respectively. Insets are results for $C_\epsilon$ and $\overline{C}$ for $U/tN=0$ with the same labels as the main figures. }\label{fig: fig1}
\end{figure}

Now we turn to the quantity $\overline{C}=\int_{-\Delta}^\Delta d\epsilon{C}_\epsilon$, which is related to the area between $\epsilon\in[-\Delta, \Delta]$ below the curve of $C_\epsilon$. When $U$ is small, averaging $C_{\epsilon}$ according to Eq.(\ref{ave}) smooths out the small bumps on the curves. The non-monotonic behavior of $\overline{C}$ is largely suppressed, as shown by the solid(brown) curve in Fig.(1B). On the other hand, we have seen that the resonance feature of the curves becomes significant for large $U$.  When $\Delta$ is not large enough, the contribution from regions under the peaks to the total area is dominant. As a result, peaks can still be clearly resolved for small values of $\Delta$, as shown by the dotted(blue) and dash-dotted(red) curves in Fig.(1B). For large $\Delta$, the peaks become wiggles on the top of the slowly decaying curves of $\overline{C}$. It is also quite clear that $\overline{C}$ first increases when the disorder strength $\Delta$ grows from zero, if $\Delta<\Delta^*$, where $\Delta^*=U$ is the location of  the first peak. This fact shows the significant difference of the disorder effect between the non-interacting and interacting systems. In the presence of interaction, the disorder effect is non-monotonic.
\begin{figure}[tbp]
\begin{center}
\includegraphics[width=1.8 in]{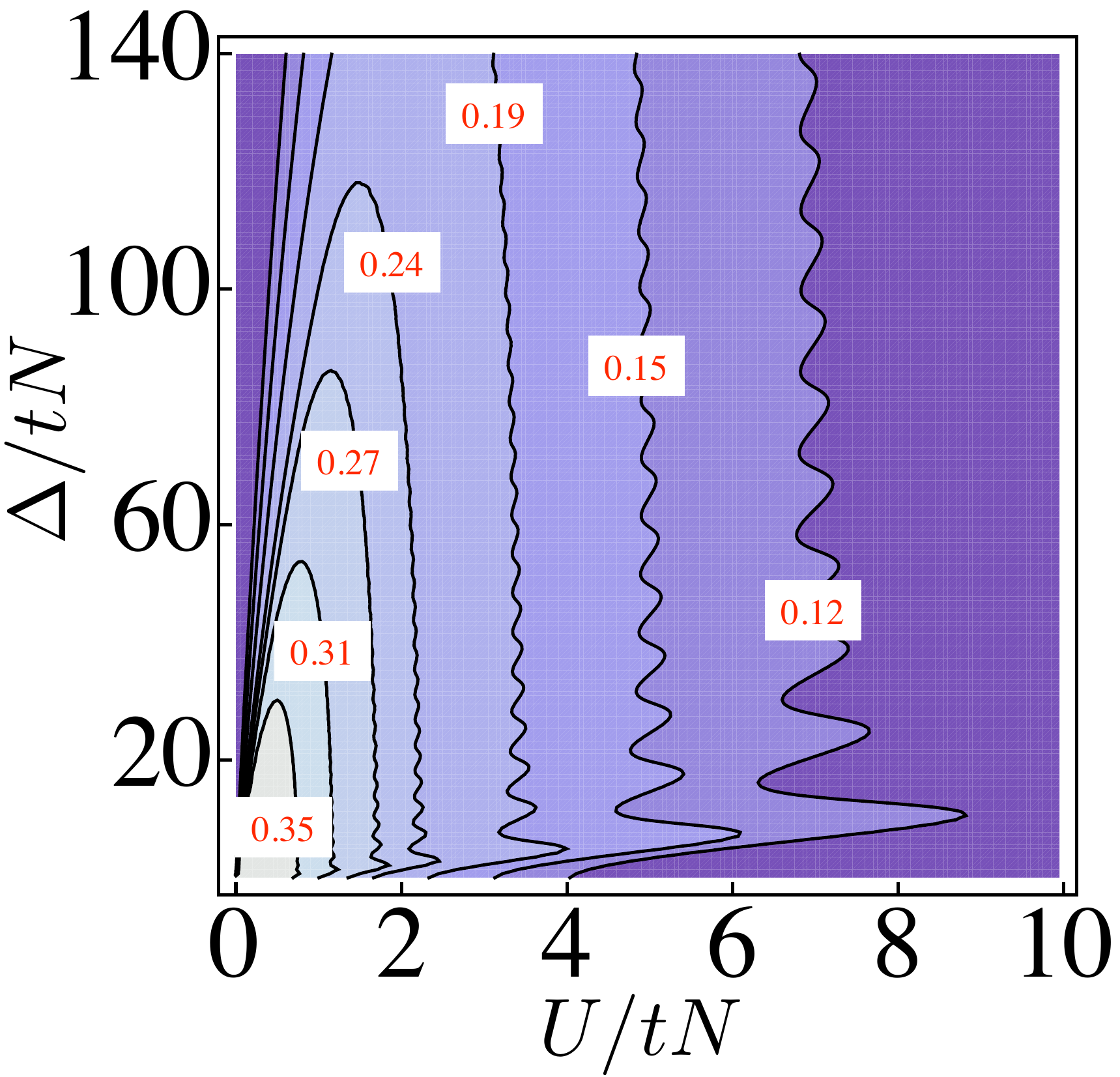}
\end{center}
\caption{A contour plot of $\overline{C}$ as a function of $\Delta/tN$ and $U/tN$ at zero temperature with the values of contours marked down.}\label{fig: fig2}
\end{figure}

It is helpful to look at the contours of $\overline{C}$ as a function of $\Delta/t$ and $U/t$ in order to obtain a complete picture. Fig.(2) shows that for large $U$, the contours first bend to the right hand side when $\Delta$ increases from zero. We note that the contours of $\overline{C}$ have wiggles at large $U$. These wiggles arise from the structures in $\overline{C}$ as discussed in the last paragraph. For a fixed value of $U$, away from the positions of the peaks, $\overline{C}$ changes slowly, corresponding to the parts of the contour which are nearly parallel to the $\Delta$ axis. When approaching the peaks, $\overline{C}$ quickly increases. As a result, the contours bend towards the $U$ axis, forming a wiggly shape. The topology of Fig.(2) is very similar to the nontrivial structure of the phase boundary obtained by recent numerical simulations for the disordered BH model at an integer filling\cite{QMC, DMRG}. In the latter case, the phase boundary can be viewed as the contour for the order parameters $\langle b_i\rangle=0$ in the thermodynamic limit. 

We emphasize that the similarity of the topology between Fig.(2) and the phase diagram of the disordered BH model is not accidental. The lattice model can be viewed as the thermodynamic limit of the two-site problem. Moreover, the physics of the interplay between disorder and interaction is the same in both cases. Interaction discretizes the energies of the Fock states in each site, and suppresses the tunneling between different sites. However, disorder introduces relative energy shifts of the Fock states at different sites. In a randomly distributed disorder potential of large enough strength, there are always possibilities for neighboring sites to have nearly degenerate Fock states. Effectively, the tunneling of the particles as well as the spatial phase coherence will then be enhanced. If the disorder strength increases further, the weight of those configurations favoring tunneling in all the configurations of the random potential decreases. The phase coherence is thus eventually suppressed by very strong disorder, as seen also in the numerical simulations of the full disordered BH model\cite{NT, RS, QMC, DMRG}. It is worthwhile to point out that a quantitative understanding of the establishment of a long range order in the disordered BH mode requires taking into account long-range correlations beyond neighboring sites\cite{fullBH}. Nevertheless, our two-site model qualitatively reveals the underlying physics for the enhancement of phase coherence through weak disorder.

{\bf The answer to Q2}: Having answered {\bf Q1}, the answer to {\bf Q2} becomes clear. Fig. (3A) shows the dependence of $\overline{C}$ on $U/t$ at fixed $\Delta/t$. For $\Delta=0$, interaction suppresses phase coherence monotonically. For $\Delta\neq0$, interaction first enhances phase coherence before suppressing it. These results are reminiscent of a similar behavior of the superfluid density as a function of $U/t$ in the full disordered BH model\cite{NT, RS, QMC}. In both cases, interaction screens the disordered potential, since some particles occupy the sites with lower on-site energies and thus smooth out the effective potential for the remaining particles. The spatial phase coherence is then enhanced. However, if interaction becomes very strong, interaction itself eventually destroys phase coherence or superfluid. 

We have so far seen an interesting ``two negatives make a positive" effect on the spatial phase coherence of a system owing to the interplay between interaction and disorder. With only disorder or interaction, the spatial phase coherence is suppressed either by the single particle localization or the emergence of Mott state. When both interaction and disorder are present, our exactly solvable model clearly shows that the spatial phase coherence is enhanced in the parameter regime where they are comparable in strength.  

\begin{figure}[tbp]
\begin{center}
\includegraphics[width=3.4 in]{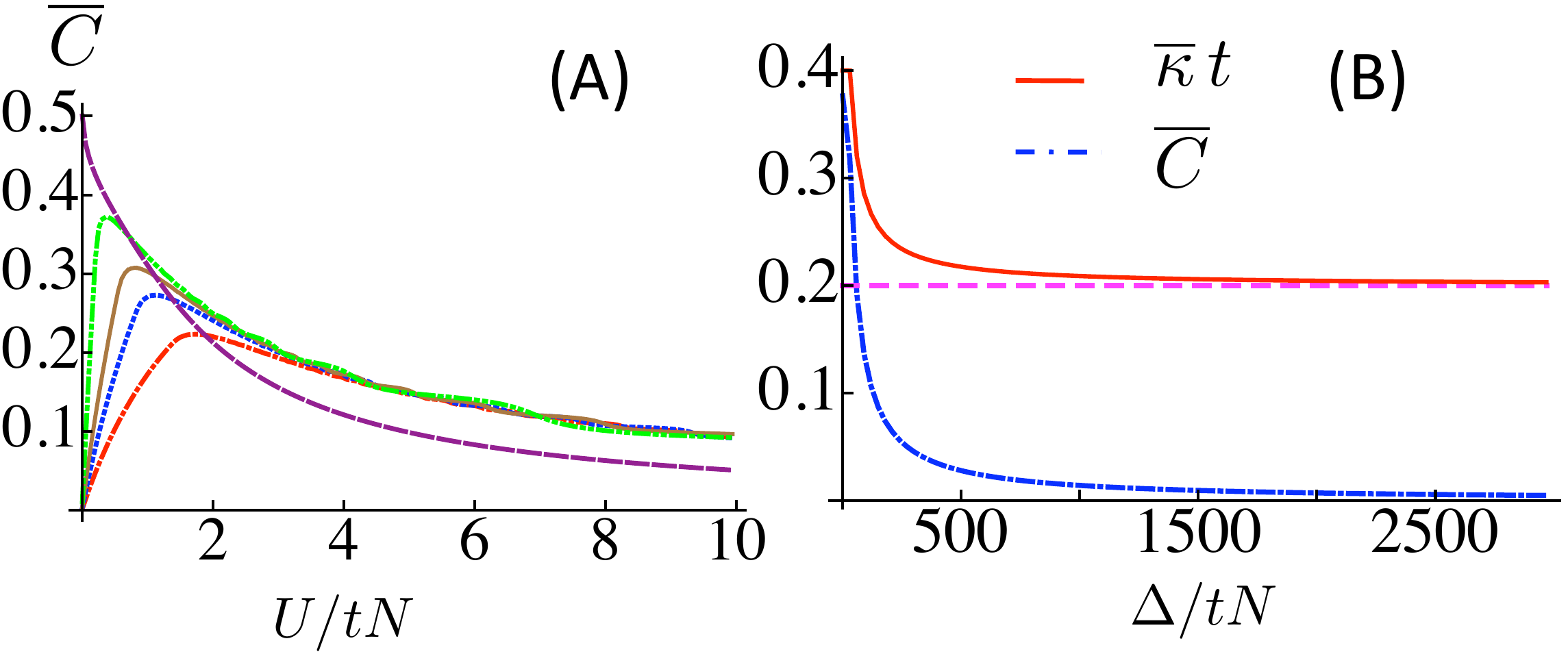}
\end{center}
\caption{Left: $\overline{C}$ as a function of $U/tN$ at $\Delta/tN=0$(dashed, purple), $20.4$(dash-double-dotted, green), $55.5$(solid, brown), $83.7$(dotted, blue) and $140$(dash-dotted, red). Right: compressibility(solid red) and $\overline{C}$ (dash-dotted blue) as a function of $\Delta/tN$ at $U/tN=0.5$. The dashed(magenta) line represents the asymptotic value of $\overline{C}$ at large interaction limit. }
\end{figure}

{\bf The answer to Q3}: To extract the compressibility $\kappa_\epsilon$, we first calculate the chemical potential, $\mu_\epsilon=(E^0_\epsilon(N+\delta N)-E^0_\epsilon(N))/\delta N$, where $E^0_\epsilon(N)$ is the ground state energy of $N$ particles at fixed $\epsilon$. From the dependence of $\mu_\epsilon$ on the particle number $N$, $\kappa_\epsilon=\partial N/\partial \mu_\epsilon$ can then be obtained. The ensemble averaged compressibility $\overline{\kappa}$ can be calculated in the same manner as $\overline{C}$, i.e., $\overline{\kappa}=\int_{-\Delta}^\Delta d\epsilon \kappa_\epsilon$. The dependence of $\overline{\kappa}$ on $\Delta$ is shown in Fig.(3B). For comparison, the result of $\overline{C}$ for the same $U$ is also shown.  

An interesting feature of the compressibility in the minimal model is that it remains constant for large disorder strength, even though the phase coherence $C$ decreases for very large disorder. In the large $\epsilon$ limit, almost all the particles fall into one of the two wells, $E^0_\epsilon\rightarrow UN(N-1)/2-\epsilon N$. It is easy to see $\kappa_\epsilon\rightarrow U^{-1}$. The ensemble averaged value $\overline{\kappa}$ for large $\Delta$ is mainly determined by the contribution from $\kappa_\epsilon$ in the large $\epsilon$ region. As a result, $\overline{\kappa}$ also approaches a constant value $U^{-1}$ when $\Delta$ is very large. We can view the state in this regime as an analog of the bose glass phase in the thermodynamic limit, which has vanishing order parameter $\langle b_i\rangle=0$(or supefluid density $\rho_s=0$) but a finite compressibility $\kappa$.  To fully understand the bose glass phase in the disordered BH model, one needs to consider correlations beyond nearest neighboring sites, which also contribute to the finite compressibility of bose glass phase, despite the absence of a long range order. Nevertheless, our two-site model provides a simple example to demonstrate why the compressibility of a disordered system can remain finite after the phase coherence has been destroyed at large disorder strength.   

\begin{figure}[tbp]
\begin{center}
\includegraphics[width=3.4in]{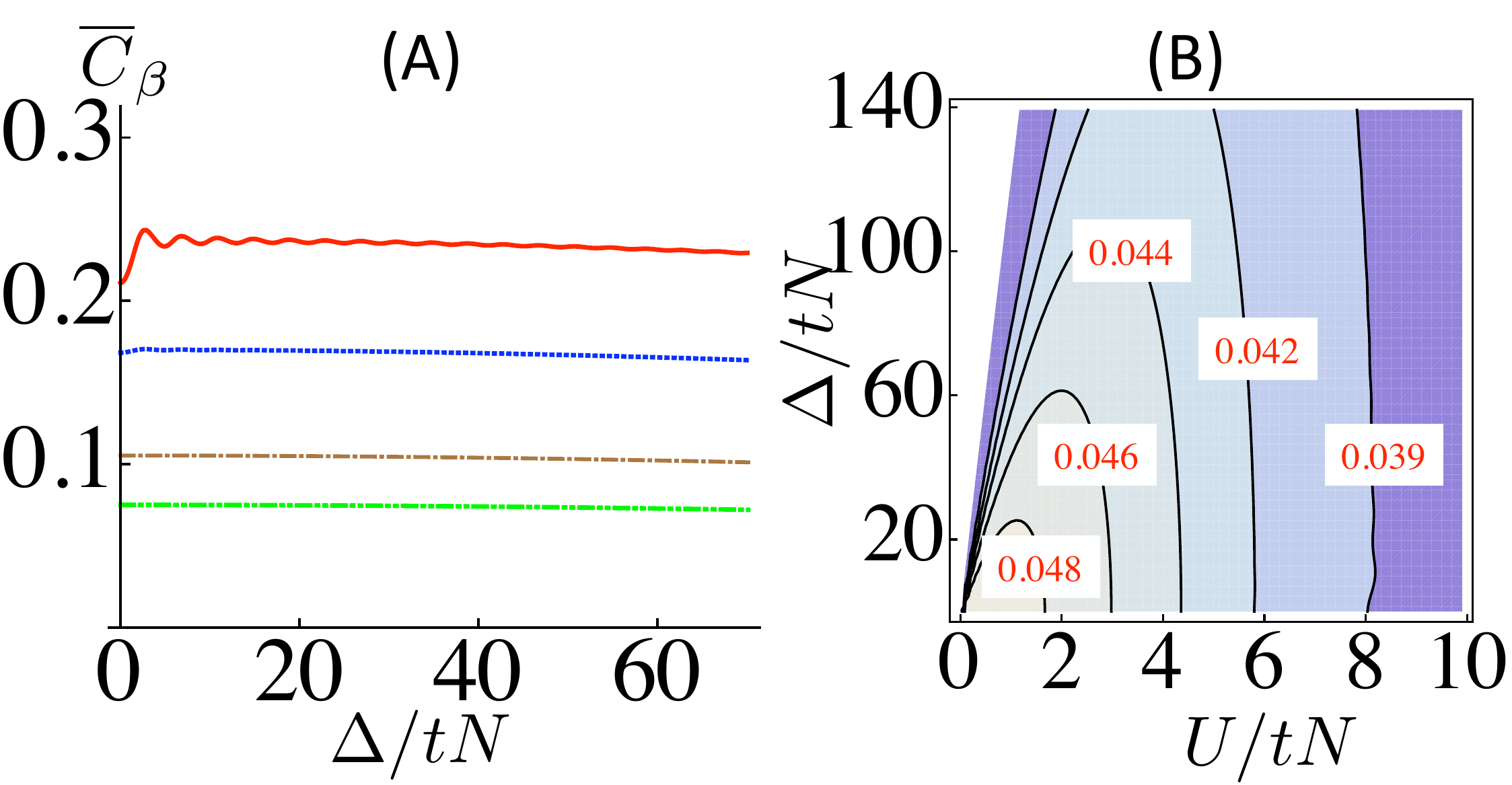}
\end{center}
\caption{Left: $\overline{C_\beta}$ as a function of $\Delta/tN$ at different temperatures for a fixed interaction $U/tN=2$. From top to bottom, $T/tN=0.4$, $1$, $2$, $3$. Right: a contour plot of $\overline{C}$ as a function of $U/tN$ and $\Delta/tN$ at the temperature $T/tN=5$. }\label{fig: fig4}
\end{figure}

{\bf The answer to Q4}: We have seen that increasing disorder at large interaction leads to an increase of phase coherence at zero temperature.  The topology of the phase diagram at zero temperature obtained from recent numerical simulations of the full disordered BH model also indicates a transition from the insulating phase to the superfluid phase with increasing disorder strength\cite{QMC,DMRG}. However, a recent experiment did not find any evidence for the increase of the condensate fraction with increasing disorder strength\cite{Demarco3}. We will show that the topology of the contours of $\overline{C}$ changes at finite temperatures, namely, the contours may bend to the left hand side when the disorder strength increases from zero. This fact may help understand why the experiment at finite temperatures did not observe disorder enhanced phase coherence. 

The value of the phase coherence at finite temperatures can be calculated by $C_{\beta,\epsilon}=(\sum_ne^{-\beta E_n}M_{l.l+1}\psi^n_{l}\psi^n_{l+1})/Z$, where the subscript $\beta$ denotes the thermal average, $\phi^n_l$  is the $n$th eigenfunction with eigenenergy $E_n$, $Z=\sum_n e^{-\beta E_n}$, and $\overline{C}_\beta=\int _{-\Delta}^\Delta d\epsilon C_{\beta,\epsilon}$. The results for $\overline{C}_\beta$ as a function of $\Delta$ for a fixed $U$ are shown in Fig.(4A) at different temperatures.  At low temperatures, the wiggles on the curve of $\overline{C}_\beta$ retain. When temperature increases,  the wiggles are gradually suppressed. At high enough temperatures, the wiggles completely vanish, leading to a monotonic decrease of $\overline{C}_\beta$. Consequently, the topology of the contours of $\overline{C}_\beta$ changes completely, as shown in Fig(4B).  Since the minimal model appears to describe the full disordered model rather well qualitatively, it is reasonable to assume that the phase diagram for the lattice model at finite temperatures will change in a similar manner as Fig.(4B). It has also pointed out that the weak superfluidity in the so called finger region, corresponding to the regions near the wiggles in Fig.(2) in our case, can be easily suppressed by finite temperature effects\cite{QMC,Hofstetter}.

As a conclusion, we have proposed and solved a simple double-well model which incorporates many key ingredients of the disordered BH model. It strikingly captures a number of novel features of the quantum phase diagram of the full BH model. The ``two negatives make a positive" effect is expected to be a general feature when the strengths of disorder and interaction are comparable. We believe that our minimal model could be a simple theoretical paradigm for understanding the details of the full disordered interacting quantum phase diagram in many situations. We expect our work will stimulate more theoretical studies to go beyond the two-site model and take into account the long range efforts in a lattice for a complete understanding of the disordered BH model.

The authors acknowledge Tin-Lun Ho for helpful discussions. This work is supported by JQI-NSF-PFC and ARO-DARPA-OLE.


\begin{thebibliography}{99}
\bibitem{fullBH}In the full disordered BH model,  with increasing disorder strength, there are either successive phase transitions in which superfluid phase is present, or only one phase transition from Mott insulator to bose glass\cite{QMC}. Whereas phase coherence is enhanced by disorder in both cases,  a long range order is established in the first case but not in the second one.
\end{thebibliography}
\end{document}